\documentclass[aps,prb,showpacs,titlepage,preprint,12pt]{revtex4}
\usepackage{graphicx}
\usepackage{amsmath}
\usepackage{amsfonts}
\usepackage{amssymb}
\begin{document}
\title{ On Superconductivity of One-Dimensional
 Channel with Strong Electron-Electron Interaction. }
\author{V.V.Afonin, S.V.Gantsevich }
\affiliation{Ioffe Physical-Technical Institute
 of the Russian Academy of Sciences, 194021, Saint Petersburg, Russia\\
vasili.afonin@mail.ioffe.ru,\quad sergei.elur@mail.ioffe.ru}
\begin{abstract}
\baselineskip=2.5ex We study the ground state of one-dimensional
channel with strong attractive  electron-electron interaction at
low temperatures. In spite of the fact that at low temperatures
the ground state of one-dimensional attracting  electrons is a
state with a macroscopically large number of cooperons, the
resulting superconducting phase has a number of significant
differences. Namely, the order parameter (which should appear in
the superconducting phase according to Landau's phenomenological
theory) turns out to be zero. However, elastic impurities
implanted in a one-dimensional channel will not lead to
dissipation of the supercurrent associated with the condensate
movement as a whole.
\end{abstract}
 \maketitle \baselineskip=2.5ex
\begin{center}
\section{Introduction, Problem Statement and Qualitative Discussion of the Results}
\end{center}
In connection with the development of nanotechnologies, which
allow for the production of one-dimensional ballistic channels,
interest in the theory describing these objects was revived.
Another reason for traditionally stimulating theoretical work in
this area is that exactly solvable one-dimensional models (see,
for example, the papers Luttinger \cite{L}, Schwinger \cite{Sh})
gave hope to understand, at the qualitative level, the phenomena
occurring in a higher dimension systems. For example, the
Schwinger model,  for a long time, has been the only field theory
to discuss the confinement problem. It should be said that, over a
long time, the Luttinger model (one-dimensional interacting
electrons without backscattering) was considered a purely
theoretical problem too. It gave a number of interesting (from the
theory point of view) results. Later, it served as the basis for
considering problems related with quasi-one-dimensional problems.
However, at the beginning of this century, the first experimental
works on quantum wires appeared. One could hope that electrons
occupy only the lowest level of transverse quantization
\cite{B,Tak}. As a result, the electronic transport in the objects
should have been described by a one-dimensional model. It
stimulated works related to the calculation of the conductance in
the one-dimensional channel, the influence of elastic impurities
implanted into the channel \cite {FK}, and the effects associated
with the screening of electron--electron interactions by a
three-dimensional environment.
\par
In this paper, we will discuss the other question. We will show
that (as was to be expected) the ground state of the Luttinger
model with a strong electron--electron attraction is a state with
a long-range order. Using direct calculation, we will make sure
that, in the case of single-component electrons, it contains a
macroscopically large number of Cooper pairs (correlated states
acting as Bose-like particles). For the multicomponent system,
correlated states consist of more complex correlated complexes:
fours for two-component fermions (electrons with spin), eights for
conducting nanotubes (four-component fermions), etc. This is a
phenomenon known in relativistic theories as the t'Hooft
principle. According to it, the interaction of $n$-component
fermions leads to an appearance in $|in>$ and $|out>$ states the
correlated complexes consisting of the maximum possible fermions
number permitted by the Pauli principle \cite{thooft}. In our
problem, this means that, for two-component fermions, the
correlated complexes consist of two electrons with opposite spins
moving right and the same electrons moving left (see Equation
(\ref{gr12mn})). We will call all such complexes cooperons. They
always consist of an even number of electrons, and their total
spin is zero. From the statistics view point, such objects look
like extensive bosons. Therefore, the long-range order arising at
the system has a purely statistical nature.
\par
Before discussing the detailed form of the ground state in the
one-dimensional case, we briefly discuss the questions of the
physical picture: the three-dimensional superconductivity that is
important for the one-dimension case. The fundamental property of
what we now call the phase with broken symmetry is the ``rigidity
of the ground-state wave function''. It was understood at the
early stages of superconductivity theory development (F.London
\cite{Lond}).
After understanding instability of the normal metal ground state
(Fermi sphere, $| F>$) relative to infinitely small attractions
between electrons, one made arguments in favor of the existence of
correlated objects consisting of two electrons with opposite
momenta  behaving like extensive boson particles in the new ground
state (Cooper's pairs \cite{BKS}). It was the basis for the
creation of the consistent theory of superconductivity.
\par
At a qualitative level, the connection between the rigidity of the
ground state wave function and the existence of a macroscopic
bosons number (i.e., the boson number increase with an increase
the system volume, $L$) was understood, apparently, by R.P.
Feynman. In essence, it is based on the normalization factor
$\sqrt{N + 1}$, arising from the action of the boson creation
operator on $N$-bosons wave function. (Let the state correspond to
the movement of all Cooper pairs in one direction---the
supercurrent). Let us ask ourselves the following question:
What is the probability of finding a boson at another state
elastically scattered by an impurity? The ratio of the
probabilities (leaves in the $N$-bosons state or scatters to the
empty one) is about $1/N$ \cite{BE} (Bose-Einstein principle). If
$ 1/ N \sim 1 / L ^ {\beta}, $ ($ 0 <\beta \le 1 $), then this
relationship is vanishingly small. This shows that the coherence
(rigidity) of the ground state requires only a macroscopically
large number of Bose particles in the state and not a finite
density of pairs. The finite density of Cooper pairs, $ \beta = 1
$, corresponds to the usual second-order phase transition: $ \beta
<1 $ is the Kosterlitz--Thoules--Berezinsky (BKT) phase
\cite{Ber,KT}.
\par
In fact, this picture needs to clarify the symmetry properties of
the scattering, which can transfer the boson to an empty state.
Symmetry considerations are extremely important for second-order
phase transitions. It was pointed out in Landau phenomenological
theory \cite{Land}. Its starting point was the introduction of an
order parameter, i.e., a quantity that would have to be zero due
to the symmetry of the Hamiltonian. However, it turns out to be
nonzero due to the fact that the symmetry of the ground state
below the transition point turns out to be lower than the symmetry
of the Hamiltonian. Let us discuss what this means in terms of
quantum mechanics. The appearance of the nonzero order parameter
signifies that a quantum mechanical operator connected with this
quantity has the nonzero matrix element $<\Omega|A|\Omega>$
despite the fact that this contradicts the Hamiltonian symmetry.
(Here, $|\Omega> $ is the ground state wave function). The
invariance of the Hamiltonian with respect to any transformation
means the existence of the conservation law and, corresponding to
it, eigenvalues and wave functions $ | \Omega> _n $. At the same
time, the operator $ {\hat A} $ has no diagonal matrix elements in
the basis. Therefore, the existence of a nonzero order parameter $
<\Omega |{\hat A} | \Omega> $ requires that the ground state wave
function should be a package $ | \Omega> = \sum _ {\triangle n}
C_n | \Omega> _n $.  In turn, this suggests that many states $ |
\Omega>_n $ with different quantum numbers should be degenerate in
energy. Exact energy degeneration of the states is needed for a
temperature ($ T $) equal to zero. For finite temperatures, the
packet can have finite breadth with the level interval of the
order of $T_d$. We will call it the degeneration temperature. (It
should be calculated in a microscopic theory.) In such a case, the
transition between states with different quantum numbers is
possible only for $ T > T_d $, and at a low temperature region $ T
\ll T_d $, the existence of a nonzero order parameter is
impossible. (This requirement is not important for
three-dimensional superconductors, where the phase transition
temperature ($ T_c $) is determined by the constants of the theory
and the distance between the levels is determined by the size
sample. Therefore, at the limit $L \to \infty$\, $ T_d \to 0 $,
while $ T_c $ remains finite. Therefore, for a three-dimensional
sample, the question is irrelevant, though the case of extremely
low temperatures may be an exclusion. However, as we will see
later, it is important in a one-dimensional case.)
\par
In the case of three-dimensional superconductivity, the ground
state ceases to be invariant with respect to the gauge
transformation ($ \psi \to \exp{(i \Lambda)} \psi $). The
Hamiltonian invariance with respect to the gauge transformation
leads to the conservation of electric charge. Therefore, in the
superconductivity problem under $ | \Omega>_n $, one should
understand states with a fixed value of electric charge and that
the operator $ \hat A $ is an operator of the observed value
changing the charge of a state. In particular, the violation of
the gauge invariance of the theory results in the fact that the
charged particles fall out to the condensate. They can consist of
two electrons or two holes with opposite momenta. In the usually
discussed temperature region, above $ T_d $ but below $ T_c $,
states with different charges have the same energy. As a result,
the ground state wave function is non-invariant with respect to
the gauge transformation, and one has a nonzero order parameter.
\par
Another matter is the low-temperature region $ T \ll T_d $. Here,
the ground state is non-degenerate in charge, i.e., an order
parameter turns out to be zero. Moreover, if the ground state has
a charge equal to zero (usually, it is the lowest energy state),
then in this region, there is no global symmetry breaking. In this
case, the ground-state wave function has the same symmetry as the
Hamiltonian despite the fact that it consists of Cooper pairs and
the wave function of each pair is non-invariant with respect to
the gauge transformation. However, the presence of impurities in
this case does not result in  resistance appearance due to the
factor $ 1/N \sim 1 /L^{\beta} $ in the scattering probability. It
is clear that our statement about suppression in the cooperon
scattering does not apply to the weak e--e interaction case
($\beta\ll 1$). It is exactly the region where a renormalization
group approach works (see Reference {\cite{Gh} and the references
there within).
\par
Returning to properties of the scattering (which is suppressed by
many Bose particles that fall out to condensate),
it should  be invariant with respect to broken symmetry.
Otherwise, scattering (it will be strong in a one-dimension case
due to e--e interaction renormalization) leads to the level shift
and can break up the whole package $|\Omega>$. The electrons
scattering on elastic impurities is invariant with respect to the
gauge transformation. Therefore, this scattering will be
suppressed by the macroscopic number of Cooper pairs existing in
the ground state.
\par
In the next section, we will calculate the ground state wave
function in the Luttinger model with a sufficiently strong
attraction \cite{rev2}. We will show that, at $ T = 0$, the ground
state of a 1-D channel contains a macroscopically large cooperons
number. However, the number of correlated complexes consisting of
electrons and holes is equal to each other. Thus, the ground state
has an electric charge equal to zero and the wave function of the
ground state is invariant with respect to the gauge
transformation. It makes the existence of the order parameter
impossible. (Note, we are talking not about the density of the
order parameter but about its total value. Normally, the latter is
proportional to the sample volume.) The phase with a global
violation of gauge invariance under these conditions is not
realized because, in the case of a strong attractive e--e
interaction, $ T_d \gg T_c $. The small value of the phase
transition temperature is easy to understand, taking into account
that, at gap-superconductor, the transition temperature is of the
order of normal excitations gap. (Careful consideration of normal
excitations problem is given in Reference \cite{F1}.) The spectrum
of these excitations is well-known and, for the one-component
electrons, is $v_c|p_n|$ (here, $ v_c(p_n)= v_f \sqrt{ 1 +V_0(p_n)
/\pi v_f}$, $v_f $ is the Fermi velocity, and $p_n$ is the
electron momentum equals to $2\pi n/L$. The quantity $V_0(p_n)$ is
the e--e interaction "potential,"\ negative in the problem. We use
the system units $ \hbar = 1 $). Thus, we see that the gap in the
excitation spectrum is of the order of $ v_c /L $. If the length
of a one-dimensional channel is of the order of a micron and the
electron concentration is about of metallic, one can give the
order of magnitude estimation for transition temperature: $ v_c /
L \sim 1^oK \cdot (v_c / v_f $) while $ T_d \sim v_f / L $ ( in
fact, it is the only remaining quantity with the dimension of
energy and is proportional to $ 1 / L$). The specifics of
one-dimensional superconductivity is in a small critical
temperature in comparison with the energy of degeneracy.
Nevertheless, the absence of a global symmetry breaking at
temperatures lower than the degeneration temperature does not
prevent the existence of a permanent supercurrent (the motion of
the condensate as a whole). In this sense, it is possible to speak
of a phase transition to the superconducting state in a
one-dimensional channel.
\par
\section{Calculation of the Ground State Wave Function}
Our task is to calculate the ground state wave function. In
principle, one can find it  by solving the Schr\"odinger equation.
However, due to a strong electron--electron interaction, the
equation actually has an infinite number of spatial variables  and
one has no effective method for solving this equation.
\par
Another object that explicitly contains all information about the
system is the well-known evolution operator. In the Schr\"odinger
representation (dependence on   time is transferred to the wave
functions), it can be represented as
\begin{equation}
S\left( \tau \right) = \sum_{m,n} |n> <n|\exp{(-i\hat H\tau)}|m> <m| .
\label{s1}
\end{equation}
The evolution operator expresses the development over a finite
time $ \tau $ as an exact (including interaction) initial state $
<m | $ to all possible final states $ | n> $
(these states have yet to be calculated).
By ``summing up'' over all states, we mean the enumeration of all
initial and final states, and the indices
$m$ and $n$
are complete description of this state. In our case,  complete
description of the state reduces to specifying the particles
number and their quantum numbers. We assume that the wave
functions of the system at the initial time $ t = 0 $ (final $ t =
\tau $) are expressed in terms of the annihilation (creation)
operators of the right (left) electrons (holes),  $\hat a(\hat
b)_{R,L}$. They are determined according to the following:
\begin{equation}
\hat \Psi_{R,L} \left( x  \right)= \int \limits_0^{\infty}\frac{dp}{2 \pi}
\left( \exp\left( \pm ipx \right)\hat a_{R,L}\left( p  \right) +
\exp\left( \mp ipx \right)\hat b_{R,L}^{\dag}\left( p  \right)\right)=
\label{psi}
\end{equation}
$$
= \hat a_{R,L}\left( x  \right) +
\hat b_{R,L}^{\dag}\left(x  \right), 
$$
and $H$
 is the Hamiltonian of the system. In the case
of one-component electrons, it is as follows:
\begin{equation}
\hat H = \int dx \left[ \hat \Psi^{\dag}_R \left( x \right) v_f\left( -i
\partial_x \right) \hat \Psi_R \left( x  \right) +
 \hat \Psi^{\dag}_L \left( x \right) v_f i\partial_x
\hat \Psi_L \left( x  \right) \right]\\   +
\int dx dy \varrho \left( x  \right) V_0\left( x-y \right)
\varrho\left( y  \right) \nonumber .
\label{g5}
\end{equation}
Here, $ \varrho $ is the electron density equal to $ \varrho =
\varrho_R + \varrho_L $. We will count all momenta ($ p $) from
Fermi one ($ p_f$). In addition, we will use periodic boundary
conditions, i.e., strictly speaking, we should consider the
momentum as discrete. However, if the condition $ p_f L \gg 1 $ is
fulfilled, in most calculationsm the sum can be replaced by an
integral (as it is written in  Equation (\ref {psi})). The
quantities $\Psi_ {R, L}$ are the wave functions of right and left
electrons. It can be determined through the complete electron wave
functions according to the following:
$$
\hat\psi (x)={\exp(ip_fx)}\hat\Psi_R(x)+{\exp(-ip_fx)}\hat\Psi_L(x)
$$
Note that, after transition to the "complex time" \ $ \tau \to -i
/ T $, the evolution operator becomes the usual density matrix.
Recall that, by the indices "$m$" and "$n$" in this density
matrix, one needs to understand not only the quantum numbers of an
electron but also the electrons number in a given state.
\par
The expression for the evolution operator can be written as a functional integral:
\begin{equation}
S(\tau) =\int_{\left( \overline{\Psi} ,\Psi\right) }
{\cal D}\Psi {\cal D} \overline{\Psi}
\exp{\left({\cal S}\right)
\left( \overline{\Psi} ,\Psi\right)}.
\label{s2}
\end{equation}
Here, $\overline{\Psi}$ and $\Psi$ are the electron fields
(Grassmann variables) and ${\cal S}$ is the action. Dependence of
the evolution integral on $ \tau $ arises from the fact that the
integration over time is performed not on the infinite interval $
(- \infty, \infty) $ but on the interval $ (0, \tau) $. However,
the main difference in the Equation (\ref {s2}) from the ordinary
Feynman partition function consists in the fact that the fields $
(\overline {\Psi}, \Psi) $ do not tend to zero at the ends of the
time integration domain
but satisfy the following boundary conditions:\\
At $t  \to +0 $,
\begin{equation}
\Psi_{R,L} \left( x,t \right) = a_{R,L}\left( x  \right) +
\mbox {arbitrary negative frequency part}
$$
$$
\overline{\Psi}_{R,L}\left( x,t \right) =
b_{R,L}\left( x  \right)+
\mbox {arbitrary negative frequency part}
\label{gpsi}
\end{equation}
At  $t  \to \tau-0$,
$$
\Psi_{R,L} \left( x,t \right) =  b^{\dag }_{R,L}\left( x  \right) +
\mbox{arbitrary positive frequency part}
$$
$$
\overline{\Psi}_{R,L}\left( x,t \right) = a^{\dag }_{R,L}\left( x
\right)+\mbox {arbitrary positive frequency part}
$$
\par
In these boundary conditions, arbitrary positive (negative)
frequency parts arise because, after acting on the Fermi sphere ($
|F> $ is the vacuum state of our theory), they give zero. Note
that the exact states (that we have to find yet) in this
representation have the form $ \phi (x_1, x_2, \cdot \cdot \cdot)
\hat a ^{\dag}_{R, L} (x_1) \hat b^{\dag}_{R, L} (x_2) \cdot \cdot
\cdot | F> $. The boundary conditions dramatically complicate the
wave function calculation in comparison with the Green function
one (the latter satisfies the zero boundary conditions). In
addition, summation over all exact initial and final states in the
expression for the evolution Operator in Equation (\ref{s1}), in
essence, means that the matrix element is the sum of all
n-particle Green functions. As we will see, the ground state wave
functions can be analytically calculated only for a sufficiently
strong e--e interaction.
\par
Now, let us consider the initial many-body state ($|F> $) over
which the creation and annihilation operators are defined. As
usual, the operators $\hat\Psi^{\dag}_{R,L}$ are defined over the
empty state $|0>$ whereas the operators $\hat
a_{R,L}^{\dag}\left(x \right)$ and $\hat b_{R,L}^{\dag}\left(x
\right)$ are defined over the filled Fermi sphere. Moreover, we
suppose that our system is neutral as a whole. Therefore, we
should introduce some sort of positive charge background (``jelly
model'') and hence redefine the electric charge of the states. In
the case the vacuum, $|F> $ is electrically neutral state. Cooper
pairs $\hat\Psi^{\dag }_{R}\hat\Psi^{\dag }_{L}|0>$; $\hat a^{\dag
}_R\left( x  \right)\hat a^{\dag }_L\left( x \right)|F>$ are the
states with the charge $2e$, while the pair $\hat b^{\dag
}_L\left( x' \right)\hat b^{\dag }_R\left( x' \right)|F> $ has the
charge $-2e$. Therefore, a two-cooperons state $\hat a^{\dag
}_L\left( x' \right)\hat a^{\dag }_R\left( x' \right)\hat b^{\dag
}_L\left( y' \right)\hat b^{\dag }_R\left( y' \right)|F>$ is
neutral. The charge redefined in such a way is directly connected
with gauge symmetry, which is usually broken during the
superconductor transition. To avoid misunderstandings, we note
that the whole electric charge of the entire system, as always,
remains. In fact, its conservation is guaranteed by the
time-independence normalizing factor of the electron's total wave
function (or, in other words, the theory is unitary), unlike
changing electric charge of one state. The last simply means a
redistribution of the charge between all states (or with a
reservoir connected to the sample).
\par
We begin consideration from the limit $ T \to 0 $. For this case,
one should transform $\tau\to -i/T$. The evolution operator for
the Luttinger model with any e--e interaction was calculated in
Reference~\cite{BKT}. However, the very cumbersome result was
analyzed there only for repulsive electron potentials. Our
attention at that time was focused on the existence of the chiral
phase. The phase was new to the solid state physics.  It is
connected with the chiral symmetry violation
($\hat\Psi_{R,L}\to{\exp (\pm i\Lambda}) \hat\Psi_{R,L}$), and the
condensate is created by exciton-like neutral pairs $\hat a^{\dag
}_R\left( x  \right)\hat b^{\dag }_L\left( x  \right)$.
\par
Now, we will consider the attraction interaction. Here, we
restrict ourselves to discuss the case point-like e--e interaction
(Gorkov's model) and equilibrium electron system. We will not
reproduce a calculation of the functional integral of Equation
(\ref{s2}) in the case of attractive e--e electrons interaction.
Here, we limit ourselves by indicating the reason why the
analytical expression for the evolution operator allows us to
write an expression for the ground-state wave function (at least
in the form of an infinite series). In the following, using the
general expressions for the ground state wave function obtained
earlier in Reference \cite{BKT}, we will analyze it for the
attractive electrons.
\par
The possibility to obtain an expression for $|\Omega >$ out of the
evolution operator is based on the fact that the transition matrix
element from the initial state $ <m| $ to the final one $ |n> $
factorizes (it is represented as a product of two functions each
of which depend only on coordinates particles at initial (final)
states). Therefore, it can be explicitly represented as
$|\Omega_n>\cdot <\Omega_m|$. Later, everything will depend on how
effectively one can sum the perturbation series for the wave
function $ |\Omega_n> $.
\par
At $ T = 0 $, the ground state wave function can be written as an infinite series:
\begin{eqnarray}
|\Omega > = \sum_{n=o}^{\infty }\frac{1}{n!}\left[ \int
\frac{dxdx'}{\pi i}\frac
{\hat a^{\dag }_R\left( x  \right) \hat b^{\dag }_R\left( x ' \right) }{ x'-x -i\delta } + \right.  \nonumber \\
+ \left. \int \frac{dydy'}{2\pi i}
\frac{ \hat a^{\dag }_L\left( y  \right) \hat b^{\dag }_L\left( y ' \right) }{ y-y' -i\delta } \right]^n \exp
{{\cal S}_{cnf}}\left(x,x',..,y,y',..  \right) |F > ,
\label{gr12}
\end{eqnarray}
In this expression  the pre-exponential factor actually is the
most general type of wave function with zero electric charge, and
the whole really informative part is contained in the form of
configuration action (${\cal S}_{cnf}\left(x,x',..,y,y',..
\right)$). It is different for each term of the series:
\begin{eqnarray}
{\cal S}_{cnf}\left( x...,x'...,y...,y'.... \right)=  \frac{\pi}{
L} \alpha\sum_{n\ne 0}\frac{1}{|p_n|} {\cal R }_f \left( -p_n
\right) {\cal R }_f \left( p_n \right), \label{gr11}
\end{eqnarray}
where $\alpha =[1-v_c/v_f]/[1+v_c/v_f]$ and
\begin{equation}
{\cal R}_f\left( p \right) = \sum_{x..;x'..;y..;y'... } \theta
\left( p \right)\left[ \exp \left( ipx \right)+... - \exp \left(
ipx' \right)-... \right] + \label{gr3}
\end{equation}
$$ \theta \left( -p \right)\left[ \exp \left( ipy \right)+... -
\exp \left( ipy' \right)-... \right] .
$$
(Here and below, we will denote the letter $x$ as the coordinates
of the right electrons, $ x ^ {'} $ - holes, and, respectively, $
y $ and $ y ^ {'} $ coordinates of the left electrons and holes
that appear in each term sums (Equation~(\ref{gr12})).
``Summation'' occurs over all particles of which the creation
operators are in the pre-exponential factor in the expression for
the ground-state wave function.) In these expressions, we can
already move from the discrete to the continuous spectrum ($  L\to
\infty $). The finite size of the system for our problem is only
important until the moment of neglecting of the terms
exponentially small in the parameters $ 2 \pi n v_{(f,\ c)} / LT
$. As a result, the channel length is included only in the
parameters of the theory related to temperatures, and all
calculations, in fact, are done as for an infinite sample.
\par
In order to transform  Equation ({\ref{gr12}), one recalls that,
according to the logarithm connectedness theorem (in statistical
physics, it is known as the first Mayer theorem \cite{Y}), the
wave function can be represented as an exponent of the connected
diagrams sum \cite{FS}, i.e., connected terms of Equation
(\ref{gr12}). Disconnected diagrams, which also exist in $
|\Omega> $, are generated by the power terms of the exponent
decomposition. This means that the expression for the ground state
wave function can be represented in an explicit analytical form in
the case when the number of connected diagrams is small.
\par
The task is extremely simplified in the case of a strong e--e
interaction. As we will see, in this limit, the condensate
consists of a macroscopically large number of point-like Cooper
pairs. They do not interact with each other. In the case of
point-like e--e interactions, the scattering of the pairs is
possible if the two identical electrons  (that make up the Cooper
pairs) are at the same point. This is excluded by the Pauli
principle. At a weaker e--e interaction,  the Cooper pair acquires
a finite radius and the Pauli principle ceases to suppress the
scattering of cooperons. However, from the expression for $ v_c $,
it is seen that, with a very large constant of the attractive e--e
interaction, the excitation spectrum becomes imaginary, i.e., the
system collapses. Physically, this means that it is necessary to
modify the potential of the e--e attraction, adding a hard
repulsive core to it. It makes calculations much more complicated.
Instead, we restrict ourselves to the case of a relatively weak
interaction ($ | V_0 | / \pi v_f \le 1 $) when the spectrum of the
excitations still remains real.
\par
In order to understand the specifics of our problem, we first
consider the maximally strong attraction $ | V_0 | / \pi v_f = 1
$, i.e., $ v_c = 0 $. We will see that this limit describes a
ground state with a nonzero concentration of noninteracting Cooper
pairs and a total electric charge equals to zero. Next, we will
take into account small corrections in the configuration action,
proportional to $ v_c $, and make sure that they transform this
state into a Kosterlitz--Thoules--Berezinsky phase \cite{BKS,Ber}.
\par
Consider the first few terms of the series in Equation
(\ref{gr12}). Because of the $ \theta $ functions included in $
{\cal R}_f \left (p \right) $, in the wave function, there are
nonzero only terms containing right and left electrons
simultaneously and the number of electron and hole operators have
to be the same. The simplest of these states is  $\hat a^{\dag
}_R\left( x  \right) \hat b^{\dag }_R\left( x' \right)\hat a^{\dag
}_L\left( y  \right) \hat b^{\dag }_L\left( y' \right)|F >$. The
contribution of this configuration to the action at $ v_c = 0 $ is
 \begin{equation}
{\cal S}_{cnf}\left( x...,x'...,y...,y'.... \right) =
-\ln{\frac{\left( x-y +i\delta \right)\left( x'-y' +i\delta
\right)}{ \left( x'-y +i\delta \right)\left( x-y' +i\delta
\right)}} \label{v3a}
\end{equation}
As a result, in the expression for the wave function of the ground
state, we obtain the following:
$$
\int\frac{dxdx'dydy'}{(2\pi i)^2}\frac{\hat a^{\dag }_R\left( x
\right) \hat b^{\dag }_R\left( x' \right)\hat a^{\dag }_L\left( y
\right) \hat b^{\dag }_L\left( y'
\right)}{(x'-x-i\delta)(y-y'-i\delta)}\frac{\left( x'-y +i\delta
\right)\left( x-y' +i\delta \right)}{ \left( x-y +i\delta
\right)\left( x'-y' +i\delta \right)}.
$$
Further analysis of a contribution will be based on the analytical
properties of the creation and annihilation operators. As follows
from Equation (\ref{psi}), the operators $a^{\dag }_L\left( y
\right)$  and $\hat b^{\dag }_L\left( y' \right)$  are analytical
in the upper half-plane and $\hat a^{\dag }_R\left( x  \right)$
$\hat b^{\dag }_R\left( x' \right)$ is analytical in the lower
one. Thus, the integrals are determined by the pole residue at the
points $ x'= y'-i\delta; x = y-i \delta $ and become the product
of two disconnected diagrams describing two noninteracting Cooper
pairs.
\begin{equation}
\int dx\hat a^{\dag }_R\left( x  \right)\hat a^{\dag }_L\left( x
\right) \int dx'\hat b^{\dag }_L\left( x' \right)\hat b^{\dag
}_R\left( x' \right)|F >. \label{c2}
\end{equation}
Let us make sure that all other diagrams that are not reducible to
the power of that one are equal to zero. Consider, for example,
the six-fermion contribution to  Equation (\ref{gr12}):
$$
\frac{\hat a^{\dag }_R\left( x  \right) \hat b^{\dag }_R\left( x'
\right)\hat a^{\dag }_R\left( x_1  \right) \hat b^{\dag }_R\left(
x'_1 \right)\hat a^{\dag }_L\left( y  \right) \hat b^{\dag
}_L\left( y'
\right)}{(x'-x-i\delta)(x'_1-x_1-i\delta)(y-y'-i\delta)}\frac{\left(
x'-y +i\delta \right)\left( x_1-y' +i\delta \right)\left( x'_1-y
+i\delta \right)\left( x-y' +i\delta \right)}{\left( x-y +i\delta
\right)\left( x_1-y +i\delta \right)\left( x'-y' +i\delta
\right)\left( x'_1-y' +i\delta \right)}
$$
This integral will be determined by the poles $x\to y\to x_1;
x'\to y' \to x'_1$ in the lower half-plane, and we get the
configuration $\hat a^{\dag }_R\left( x  \right)\hat a^{\dag
}_R\left( x  \right)\hat a^{\dag }_L\left( x  \right)\cdot \hat
b^{\dag }_R\left( x' \right)\hat b^{\dag }_R\left( x' \right)\hat
b^{\dag }_L\left( x' \right)$. It is  equal to zero according to
the Pauli principle. It can be verified that the remaining
diagrams reduce to these two cases. Therefore, one has only two
connected diagrams. However, applying the logarithm connectedness
theorem, one should keep in mind that, in our case, only states
with a total electric charge equal to zero exist and that the
remaining states should be omitted. We denote as $P(Q=0)$, the
projector onto this state is as follows:
\begin{eqnarray}
|\Omega > =N_0P(Q=0)\exp \left[\int dx\hat a^{\dag }_R\left( x
\right)\hat a^{\dag }_L\left( x  \right) +\int dx'\hat b^{\dag
}_L\left( x' \right)\hat b^{\dag }_R\left( x' \right)\right]|F>,
\label{gr12f}
\end{eqnarray}
\par
Here, the $N_0$ is the normalization factor, and it can be
calculated. Thus, in this approximation, we have obtained the
condensate composed of the non-interacting Cooper pairs in one
state: $\hat a^{\dag }_R\left( x \right)\hat a^{\dag }_L\left( x
\right);\,\hat b^{\dag }_L\left( x' \right)\hat b^{\dag }_R\left(
x' \right)$  (with a nonzero density and full charge equal to
zero). We note once again the peculiarity of the obtained ground
state. On the one hand, there is no global violation of gauge
invariance (each configuration contains an equal number of
electrons and holes, and the phase of the gauge transformation
vanish). As a result,  the long range order parameter (one can
enter it, for example, like this: $< \Omega |\hat a^{\dag
}_R\left( x \right)\hat a^{\dag }_L\left( x \right)| \Omega >)$ is
equal to zero due to the electric charge conservation law. On the
other hand, one has a macroscopically large number of Bose-like
particles in one state, i.e., the system turns out to be
completely statistically correlated. (All electrons are pairing.
In the ground state, there are only Cooper pairs and their number
is macroscopically large). In this case, the Bose--Einstein
principle guarantees the impossibility of scattering pairs on
elastic impurities added to the system under consideration.
\par
We now discuss how this picture changes when small corrections
proportional to $ v_c/ v_f $ are taken into account (that is,
$\alpha $ is close to 1). In order to show that, in this case, one
has a condensate that consists of the macroscopically large number
of Cooper pairs, we should extract the correlated complexes from
the whole wave function (Equation (\ref {gr12}).) For this, one
should present the four-particles state that we discussed earlier,
$$
\int\frac{dxdx'dydy'}{(2\pi i)^2}\frac{\hat a^{\dag }_R\left( x
\right) \hat b^{\dag }_R\left( x' \right)\hat a^{\dag }_L\left( y
\right) \hat b^{\dag }_L\left( y'
\right)}{(x'-x-i\delta)(y-y'-i\delta)}[\frac{\left( x'-y +i\delta
\right)\left( x-y' +i\delta \right)}{ \left( x-y +i\delta
\right)\left( x'-y' +i\delta \right)}]^\alpha |F>
$$,
as a two correlated complexes product (their sizes should be small
compared to the sample size). They are separated from each other
by a large distance (of the order of $L$). It can be seen from
this expression  that the probability of finding the right
electron near the left hole ($|x-y'| \to 0 $) is suppressed by the
interaction and that the probability of finding the left electron
near the right one ($|x-y| \to 0 $) increases. Therefore, the
contribution from the correlated complexes consisting of right and
left electrons is determined by the cut at the points $x-y+i\delta
= 0;\,\,x'-y'+i\delta = 0 $. Moreover, we should assume that
$|y-x'| \sim |x-y'| \sim R $ $\to \infty .$  Therefore, the
contribution from the first cut is proportional to
$$
\left( 1-e^{2\pi i \alpha} \right)\int^x_{-\infty}\frac{dy}{ 2\pi
i}\frac{\hat a^{\dag }_L\left( y  \right)
}{\left(x-y\right)^{\alpha}}\frac{\left(x'-y
\right)^{\alpha}}{\left(y-y' \right)}\sim {\hat a}^{\dag
}_L(x)(\frac{d}{ R})^{1-\alpha}.$$
\par
Here, we have taken into account that the integrand converges well
for $y\to\infty$ (due to the boundary conditions, the fields
${\hat a}^{\dag }_L(y)$ are zero on the sample's surface). In this
case, the contribution from the complex is determined by the upper
limit of integration and should be cut off at $|y-x| \sim d$.
(Here, $ d $ is the ultraviolet cutoff. It is of the order of the
channel thickness: On this scale, the e--e interaction becomes
three-dimensional and one-dimensional effects are suppressed).
Similarly, the ``integration"\ over $ x'$ is done. (Later, this
procedure will be done more carefully for multicomponent
fermions.) As a result, the whole four-particles contribution
describes a state with two interacting complexes (two Cooper pairs
$ {\hat a} ^ {\dag}_R (x) {\hat a} ^ {\dag}_L (x) $ and $\hat
b^{\dag }_L\left( y  \right)\hat b^{\dag }_R$) are separated from
each other by a large distance (of the order of the channel
length):
\begin{equation}
|\Omega_{cpl}>=\int dxdy   \left( \frac{ d}{ |x-y| }
\right)^{2\left( 1- \alpha \right)} \hat a^{\dag }_R \left( x
\right) \hat a^{\dag }_L \left( x  \right)\hat b^{\dag }_L\left( y
\right)\hat b^{\dag }_R\left(y  \right)|F>. \label{rv9}
\end{equation}
Thus, instead of a disconnected diagram (\ref{c2}), at $ v_c \ne 0
$, we have a united complex consisting of two interacting Cooper
pairs. From this calculation, it becomes clear that the connected
diagrams number is infinite and that the theory does not have a
parameter that allows one to discard a large number of interacting
many-body complexes. It does not allow one to write the wave
function of the ground state in a simple form. However, the
general form of $ S_{cnf} $ (see  Equation (\ref{gr3})) is valid
for any value of $ v_c \ne 0 $. It guaranties us that all these
terms contain the same number of electron and hole operators
(i.e., their electric charge is zero). Consequently, as in the
term we considered (Equation (\ref{rv9})), all correlated
complexes consist of Cooper pairs  interacting with each other and
located at a distance of the order of $ L $. Therefore, for any $
v_c $, the ground state will consist of an infinite number of
Cooper pairs and, for $ T = 0 $, will have an electric charge
equal to zero.
\par
In fact, even the index in the expression of Equation (\ref{rv9})
is calculated up to a factor about unity. We calculated it in an
approximation in which the cooperons interact only ``directly"\
(their interaction with each other through the other pairs in the
intermediate states is not taken into account). However, it is not
difficult to correct. For that, it is necessary to renormalize the
interaction of the two labeled pairs, taking into account all
other cooperons. To do this, it suffices to calculate the
two-particle {correlator}
$<\Psi_R(x)\Psi^{\dag}_R(y)\Psi_L(x)\Psi^{\dag}_L(y)>=
<G_R(x-y)G_L(x-y)>$ at $|x-y|=R\to L  $ and to compare the degree
of $ R $.  Probably, direct renormalization of the interaction in
the wave function is possible. It requires derivation of a closed
equation for the renormalized interaction of two Cooper pairs in
the absence of a parameter. However, the final answer at the
output of this procedure is clear from the correlator. In it, the
renormalization of the interaction for the two-particles Green
function is taken into account exactly. Therefore, we should
replace $\alpha$ by
$$ \alpha_T =1-\frac{ v_c}{v_f}. $$ (Note
that, for small $ v_c $, $ \alpha_T $ differs from $ \alpha $ only
by a factor before $ v_c / v_f $.) However, even a decomposition
of the ground state wave function with respect to the interacting
pairs number  makes it possible to verify that the number of
cooperons in the ground state increases with increasing $ L $.
Indeed, it is clear from Equation (\ref{rv9}) that the probability
($ Z (R) $) to find two Cooper pairs separated by a distance $ R $
can be estimated as $\left(  1/ |R| \right)^{2\left( 1-
\alpha_T\right)}$. Therefore, the number of Cooper pairs for any
$v_c/v_f$ can be estimated from
\begin{equation}
N^2\sim \int_0^L dx dy Z(x-y)\sim L^{2\alpha_T},
\label{rv10}
\end{equation}
and $N$ grows with sample size as
$$N\sim L^{1-v_c/v_f}. $$
As we have already discussed in the Introduction, this fact alone
is sufficient to state that, according to the Bose--Einstein
principle, a supercurrent in one-dimensional channel does not
dissipate due to the scattering of the Cooper pairs on elastic
impurities. An exception can be the case of a  relatively weak
e--e interaction  where the factor $1/N$ is not too small
\cite{BE}.
\par
We now turn to the discussion of the finite temperature region $
T_d \gg T \gg T_c $. For the theory with a repulsion e--e
interaction (where the opposite inequalities are realized), the
finite temperatures were carefully studied in Reference
{\cite{BKT}}. We will not reproduce this derivation for attracting
electrons because, first, the value of $ T_c $ is understandable
from general considerations (it was discussed in Introduction).
Secondly, the impossibility existence of a phase with a global
violation gauge symmetry in this case is visible without
computing. Therefore, we confine ourselves to discuss changes in
the expression for the condensate wave function of Equation
(\ref{gr12}) in the temperature region.
\par
Dependence of the ground state wave function occurring from
boundary conditions  (from the terms with creation and
annihilation operators in  Equation (\ref{gpsi})) can be extracted
easily from the entire action. In Reference \cite{BKT}, it is
shown  that the contribution to the action from the nonzero
boundary conditions is equal to
\begin{eqnarray}
{\cal S}_0 = \sum_{i=R,L} \int dx dx'  \left[ b_i \left( x ' \right) G_i\left( x',0;x,
\epsilon \right)
 a_i\left( x  \right) +
 a^{\dag }_i\left( x ' \right) G_i\left( x',\tau;x,\tau-\epsilon\right)
 b^{\dag }_i
\left( x  \right) \right.\nonumber\\
-\left.  a_i^{\dag}\left( x ' \right) G_i\left( x',\tau;x,0\right)
a_i\left( x  \right) -
b_i\left( x ' \right) G_i\left( x',0;x,\tau\right) b_i^{\dag }
\left( x  \right) \right] .
\label{s8}
\end{eqnarray}
Therefore, $\hat S(\tau )\sim {\exp(S_0)}|F><F|$. (In the
expression for the ground state wave function $|\Omega> $
(Equation (\ref{gr12})), the factor obtained by factorizing this
contribution in the density matrix was expanded into the series.)
To get this expression, we used the Hubbard trick \cite{Hub} and
reduced the e--e interaction problem to the problem of
noninteracting electrons placed in a slowly varying external
field. (Later, a result should be averaged over all realizations
of the random Hubbard's field). Therefore, $ G_i \left( x',t';x,t
\right)$ appearing in the equation is the noninteracting electron
Green function in the external field ($ \epsilon $ in  Equation
(\ref{s8}) is a infinitely small value). It is well-known that the
Green function of an electron in an external field is proportional
to the free Green's function, which we should write in the
final-size volume:
\begin{equation}
G_{R,L}^0\left( x',t';x,t \right)=\frac{1}{L}\sum_{n\ne 0}
e^{2\pi in[x-x' \mp v_f(t-t') ]/L}.
\label{So}
\end{equation}
In order to present it in a usual pole form the neighboring terms
in the sum, it should not change much. This is so, in the first
two terms of the action $ S_0 $, the Green function enters at
coinciding times. In the third and fourth terms, the time argument
is large and, after transition to imaginary time, has the exponent
index $ -2\pi T_dn/T $. Therefore, in the region of interest to us
($T_d / T \gg 1 $), these contributions to the action are
exponentially small. That is why they were omitted at $ T \to 0 $.
Now, we have to ignore them too. The first two terms at $ T \to 0
$  have been taken into account. We have seen that they brought
about to the states with zero electric charge. In the case of
repulsive electrons (where $ T_c/T_d \gg 1 $), we could consider
the region $ T_d /T \ll 1 $. Then, the last two terms brought
about the states with nonzero chiral charges.
\par
In the temperature range $ T \gg T_c $, dramatic changes also
occur in the expression for the configuration action arising from
the first two terms. The coefficient $ \alpha $ in  Equation
(\ref{gr11}) changes. It is replaced by a function proportional to
$ |p_n|v_c/ T$. The last leads to the destruction of the
logarithmic divergence in the region of small momenta. It makes
the existence of a correlated phase impossible. Therefore, $T_c$
really is the critical temperature.
\par
An estimation of the phase transition temperature can be
understood from another point of view. As it is known, according
to the Landau--Mermin--Wagner theorem, the existence of the BKT
phase in one-dimensional systems is impossible at a nonzero
temperature. The proofs of this statement for different problems
were given first in Reference \cite{LandII,MW}.  However, this
statement relates to infinite samples and is based on the
Goldstone theorem, according to which the spontaneous violation of
continuous symmetry must be accompanied by the appearance of a
massless boson field (an acoustical mode). Such fields at finite
temperatures strongly fluctuate the results in the exponential
decay of a correlator at large distances. In our case, the
massless Goldstone's mode will be the phase of operator $
\Psi^{\dag}_R(x) \Psi^{\dag}_L(x)$, while the correlator which
characterizes the system is
$<\Psi^{\dag}_R(x)\Psi^{\dag}_L(x)\Psi_L(y)\Psi_R(y)>$ . This
correlator for an infinite sample can be calculated (see, e.g.,
Reference \cite{BKT}). The power-like behaviour of the correlator
is valid in the region $|x-y|\ll v_c/T$, and it is exponential at
the inverse limiting case. For the finite sample at temperatures
$T\ll T_c=v_c/L$ and $|x-y|\sim L$, the exponential asymptotic
does not realize inside a one-dimensional channel. Therefore,
inside of the sample, the decay of the correlator is  power-like
and slower as compared to the case of noninteracting electrons.
Just because of this, the number of cooperons in a condensate
becomes macroscopically large (see  Equation (\ref{rv10})). In
this case, the phase transition temperature can be obtained as the
estimate $|x-y|\sim L\sim  v_c/T$. Thus, at a temperature lower
than $T_c$, the BKT phase can be realized in the finite
one-dimensional sample. Actually, we only define more accurately
what one should understand under zero temperature in the
Landau--Mermin--Wagner theorem.
\par
Just in case, let us discuss our results, taking into account the
duality requirement of the theories with  attractive and repulsion
electrons. It is clearly visible in the boson representation and
requires the replacement of $ v_c / v_f <-> v_f / v_c $ (The ratio
$ v_f / v_c $ is usually called the Luttinger parameter, $ K $.)
According to it, the solution number of a hypothetically solved
many-electron Schr\"odinger equation for these problems has to be
the same. At the same time, in the theory with repulsion
electrons, there are  two correlated phases: with a global
violation of chiral symmetry and with a chiral charge equal to
zero while, in the case of attraction, there is only one:
electrically neutral. Therefore, the question arises whether this
fact does not contradict to the duality requirement? One has a new
value with the dimension of energy namely the temperature. In
fact, we have just seen that, at the level of the Hamiltonian
solutions, an oscillating  wave function with a nonzero electric
charge also exists for attractive electrons, as duality requires
(see the last two terms in Equation (\ref{s8})). Any correlated
phase will be destroyed since, at a sufficiently high temperature,
the normal excitations begin to emerge  in a large number.
Therefore, the temperature of $ T_c $ in the gap-superconductors
is determined by a gap in the spectrum of the excitations: $
\omega (p_n) = v_c | p_n | $. It follows that a distance up to an
excited state in the Luttinger model is about $ 2 \pi v_c / L $
for the both types interactions. This value plays role of a gap.
(It is clear, at least from dimension considerations, that the
replacement $ v_c -> 1 / v_c $ in the spectrum of excitations
during the transition from one problem to another is impossible.)
Therefore, a solution with a global symmetry violation is realized
in the case of repulsive electrons ($ v_f / L \ll T_c $ and terms
with the factor $ {\ exp (- | p_n | v_f / LT)} $ at the region $
T_d \ll T \ll T_c $ should be considered). However, this solution
cannot realize  in the problem with attracting electrons, since in
this case, the destruction of the superconducting phase occurs
first.
\par
We now briefly discuss the problem of multicomponent fermions.
(For repulsive electrons, this problem was discussed in Reference
\cite{Spin}). From a technical point of view, this problem is much
more cumbersome. There is no parameter under which the correlated
complexes would not dissipate each other. The latter follows from
the fact that the Pauli principle does not forbid the interaction
of the complex containing the  $\hat a^{\dag}_{R,\uparrow}$
particle with the complex having  $\hat  a^{\dag}_{R,\downarrow}$
even in the case of a point-like e--e interaction (here,
$\uparrow,\downarrow$ is a spin index; we will denote it later as
$ \alpha = \pm 1 $). (In fact, the scattering of complexes is much
stronger because, even in the case of the strongest e--e
interaction ($ v_c = 0 $), the correlated complexes are not
point-like). However, in  the problem of multicomponent fermions,
one can write an analytical expression for the ground state wave
functions, then make sure that the ground state of the system
contains the macroscopic number of Bose particles, and identify
the quantum numbers of this state.
\par
In order not to complicate the discussion, we consider the case
when the interaction of all electrons is the same. Then, the
spectrum of quasiparticle excitations for all possible states will
be also the same  and equal to
\begin{equation}
\omega (p_m) = |p_m|v_F \sqrt{ 1 +\frac{nV_0\left( p_m)  \right)}{\pi v_F}},
\label{ext}
\end{equation}
\par
Here, $ n $ is the fermion components number (we will discuss
mainly the case $ n = 2 $). The contribution of boundary
conditions  to the action remains the same (Equation (\ref {s8}))
with one exception: One should add the sum over  the spin index $
\alpha $ (In our case, each term in $ S_0 $ is diagonal over the
spin-index because the Luttinger model contains only a
density--density interaction. That is, it does not describe the
Kondo effect.) This immediately implies that, in the case of
multicomponent fermions, only the state with zero charges will be
realized (in this case, one more quantum number is added to the
electric charge---the total spin of the state). This follows from
the fact that, in $ S_0 $, the first two terms consist of the same
number of electron and hole operators (i.e., they are neutral in
the sense of any charge) and only the last two terms can lead to
configurations with nonzero quantum numbers.
The latter will bring into the action the summands smallness $
{\exp(-T_d/T_c)}$.
Therefore, the superconducting phase exists only in the region $ T \ll T_c\ll T_d $.
\par
We will now discuss the question about the form of correlated
complexes in a problem with multicomponent fermions at $ v_c = 0
$. The configuration action in this case undergoes a minimal
change; the expression in Equation (\ref{gr11}) gets the common
factor $ 1 / n $. As we will see, such interaction weakening  is
compensated by an increase in the number of particles at each
correlated complex.
\par
In order to verify this, we  first need to discuss the following
question: How, in the case of multicomponent electrons, will we
see a combination of particle operators is a correlated complex?
In the one-component fermions problem and $ v_c = 0 $, this
question was irrelevant. The infrared divergence of the action led
to the appearance of logarithmic terms of Equation (\ref{v3a}) in
$ S_{cnf} $ with a coefficient equal to 1. This meant the
emergence of new poles and the destruction of the poles arising
from the free Green function. As a result, the analytical
properties of the annihilation operators allowed us to
``calculate'' the integral, knowing the residue at the appeared
pole. After this, the expression entering the wave function of the
ground state was factorized, and for $ v_c = 0 $, we obtained
noninteracting Cooper pairs. Now, the logarithmic terms in the
action have a coefficient of $ 1/2 $  and cuts appear in the
integrand. This makes representing an operator expression in a
compact form  impossible. From a common point of view, this means
the nonlocality of cooperon  (even in the strongest interaction
case.)
\par
In order to present a general expression for the wave function of
the ground state as  interacting complexes, we should take several
steps. First of all, we should take the connected diagram in the
expression for the wave function and try to divide it on the
connected compact complexes consisting of a smaller number of
particles. To do this, we should move apart  the complexes in a
large distance ($ \sim L $), keeping the distances between the
particles inside each complex small. If, in this limit, the
expression for our contribution to $ |\Omega> $ is factorized and
c-number function in the expression  will not tend to zero, then
we can consider these complexes as "new particles."\ (In this
case, the probability to find one complex does not depend on the
coordinates of the other complex.) In other words, we should
consider our contribution to $ |\Omega> $
$$ \int
\frac{dx_+dy_+}{2\pi i}\dots\hat a^{\dag }_{R,+}\left( x_+
\right)\dots K\left(
 x_+,\dots,y_+\dots\right) a^{\dag }_{L,+}\left( y_+  \right)\dots\,.
$$
and to look at the behavior of amplitude $K\left(
 x_+,\dots,y_+\dots\right)$ provided that the coordinate difference
inside the group of variables $ x $ and $ y $ is small ($ \ll L $)
while the  coordinates difference between the variables $ x $ and
$ y $ is about $  L$. If we chose  cooperons  correctly,  then in
this limit, $K\left( x_+,\dots,y_+\dots\right)\to k_1\left(
 x_+,\dots\right)k_2\left(  y_+\dots\right)$.
This means that, as a result of the interaction, two correlated
complexes appeared and their contribution to the ground-state wave
function is as follows:
 $$
\int \frac{dx_+ }{ \sqrt{2\pi i}}\dots k_2\left(
x_+\dots,\right)\hat a^{\dag}_{R,+}\left(x_+\right)\dots
\int \frac{dy_+ }{\sqrt{2\pi i}} k_1\left( y_+\dots,\right)\hat
a^{\dag }_{L,+}\left( y_+ \right)\dots |F> ,
$$
while the connected part, $ K-k_1k_2 $, is the scattering
amplitude. (It tends to zero when $ |x_i-y_j| \to \infty $.) The
amplitude should be taken into account at renormalizing
interaction between two complexes (as it was done early when $
\alpha$ was replaced by $ \alpha_T $ at $ |\Omega_{cpl}> $). It is
reasonable to leave it in the Hamiltonian for the  interaction
cooperons  and to consider the obtained ground state as the $
|out> $ state. (It is exactly the out-state wave function of the
scattering problem). The logarithm connectedness theorem
\cite{Y,FS} ensures that the entire wave function will be
represented as an exponent of the sum connected complexes.
However, later, we will have to take into account the quantum
numbers selection rules (for our case, apply  a projector onto the
state with charge zero to the wave function).
\par
We will show how this procedure can be implemented in the case of
two-component fermions with $ v_c = 0 $. Based on previous
experience, one would expect that the smallest of the possible
complexes are ordinary Cooper pairs. The procedure described above
gives the following:
$$
\int dxdy\hat a^{\dag }_{R,+}\left( x  \right)\hat a^{\dag }_{L,-}\left( x  \right)
\hat b^{\dag }_{L,-}\left( y \right)
  \hat b^{\dag }_{R,+}
\left(  y \right)\frac{d}{ |x-y|}|F>.
$$
Therefore, their  feasible contribution to the $ |out> $ state  is
of the order of $ d / L $. (Moreover, the term does not
factorize.)
\par
The first complexes correlating at the scale of about $L$  consist
of four particles. (As it should be expected from the t'Hooft
principle  \cite{thooft}). They are derived from the term
$$
\frac{\hat a^{\dag}_{R,+}(x^{}_+)\hat
b^{\dag}_{R,+}(x^{'}_+)}{x'_+-x_+-i\delta}\frac{\hat
a^{\dag}_{R,-}(x^{}_-)\hat
b^{\dag}_{R,-}(x^{'}_-)}{x'_--x_--i\delta}\frac{\hat
a^{\dag}_{L,+}(y^{}_+)\hat
b^{\dag}_{L,+}(y^{'}_+)}{y_+-y'_+-i\delta}\frac
{a^{\dag}_{L,-}(y^{}_-)b^{\dag}_{L,-}(y^{'}_-)}{y_--y'_--i\delta }
$$
obtained as a decomposition of $ e^{S_0} $. The amplitude of $ K $
resulting from the e--e interaction and valid on all scales equals
\begin{eqnarray}
K\left( x_+,\dots  \right)=\frac{ 1 }{ x'_+-x _+-i\delta }
\frac{1 }{ x'_- -x _- -i\delta } \frac {1} { y_+-y'_+ -i\delta }
\frac{ 1 }{y_- -y'_- -i\delta }\nonumber\\
\times \frac{\sqrt {\prod_{\alpha ,\beta}
\left( x'_\alpha-y_\beta +i\delta \right) \left(
x_{\alpha}-y'_{\beta} +i\delta \right)}} { \sqrt {\prod_{
\alpha ,\beta }\left(
x_\alpha-y_\beta +i\delta \right) \left( x'_{\alpha}-y'_{\beta
}+i\delta \right)}}. \label{s10a}
\end{eqnarray}
We should select correlated complexes from our common operator
expression. The new complex appearance is due to the fact that the
interaction suppresses the probability to find an electron near a
hole (see the pole terms) and increases the probability to find
two electrons with opposite spins  close to each other (or two
holes). Therefore, the candidates for cooperons are the states
$\hat a^{\dag}_{R,+}(x_+)\hat a^{\dag}_{L,-}(y_-)\hat
a^{\dag}_{R,-}(x_-) \hat a^{\dag}_{L,+}(y_+)$ and $\hat
b^{\dag}_{R,+}(x'_+)\hat b^{\dag}_{L,-}(y'_-)\hat
b^{\dag}_{R,-}(x'_-)\hat b^{\dag}_{L,+}(y'_+)$. We should assume
that the distances between the particles entering each state are
close to each other and that the distances between these states
are large (of the order of $ L $). In this case, the amplitude
(Equation (\ref{s10a})) factorizes but does not tend to zero. Each
particle configuration enters the ground state wave function with
the factor $ V_4 $. It depends only on one-complex coordinates and
is equal to
$$
 V_4(x_{\alpha},.. )=1/ \sqrt { \prod_{
\alpha ,\beta = \pm}\left(
x_\alpha-y_\beta +i\delta \right).
}
$$
It is easy to verify that the connected part of the scattering
amplitude is not factorized and tends to zero when cooperons move
apart at a distance about $ L $:
$$
V_{coll}(x_{\alpha}.., x'_{\alpha}...)= K\left( x_+,\dots
x'_{\alpha}\dots  \right)- V_4(x_{\alpha},.. )V_4(x'_{\alpha},..
).
$$
Therefore, it should be interpreted as a vertex in the interacting
cooperons Hamiltonian. The wave function of the ground state  is
\begin{eqnarray}
| \Omega > =N_0P(Q=0)\exp [\int dx_+...V_4(x_+,..)\hat a^{\dag}_{R,+}(x_+)\hat a^{\dag}_{L,-}
(y_-)\hat a^{\dag}_{R,-}(x_-)\hat a^{\dag}_{L,+}(y_+)
+
\label{gr12mn}
\end{eqnarray}
$$+
\int dx'_+...V_4(x'_+,..)\hat b^{\dag}_{R,+}(x'_+)\hat b^{\dag}_{L,-}(y'_-)\hat b^{\dag}
_{R,-}(x'_-)\hat b^{\dag}_{L,+}(y'_+)]|F>.
$$
The state is realized when the interaction between cooperons,
$V_{coll}$, is adiabatically turned off at $t\to (\tau ,0)$, i.e.,
it is the $|out>$ state of the system in the cooperons
representation.
\par
The ground state of a strong interacting two-component electrons
system is a macroscopically large number of Bose-like strongly
correlated cooperons consisting of two right and two left
electrons (or holes) with opposite spins. The wave function of
each complex is non-invariant with respect to the gauge
transformation, but the number of complexes consisting of
electrons and holes is always equal to each other. Therefore,
there is no global violation of gauge invariance in the system.
\par
As we have just seen, even in the case of the strongest e--e
interaction ($ v_c = 0 $), our task has been reduced to the
interacting cooperons system. An analysis of diagrams describing
the mutual scattering of cooperons shows that, although their
interaction is not weak, the qualitative picture of the phenomenon
will not change. (It also happened  in the case of one-component
fermions at $ v_c \ne 0 $; see  Equation (\ref{rv9})). All that is
possible to write in this case  is the infinite decomposition of
the exact wave function cooperons by their number. This expansion
looks rather cumbersome and, from a physical point of view, does
not provide new information.
\par
Accounting the finiteness of $ v_c $ in our case leads to the
replacement of the powers in the expressions for $ V_4 $ and $ V_
{coll} $. The degrees $ 1/2 $ are replaced by $ \alpha/2
=(1/2)\cdot [1 - v_c / v_f] / [1+v_c / v_f] $. After replacement,
Equation (\ref{gr12mn}) remains correct. It is clear that the case
of $ v_c / v_f \to 1 $ is outside of our approach. In this region,
the interaction of the cooperons will lead to the destruction of
the coherence phase. In this case, it is more rational to use the
renormalization group approach formulated in the  original
electron wave functions representation \cite {Gh}.
\par
The conducting carbon nanotubes give one more example of
multicomponent interacting electrons ($n=4$). According to
Reference \cite{Aj}, a conductive nanotube can be consider  a
cylinder of small radius obtained by gluing a monatomic layer of
graphite. If the technology of nanotube production is sufficiently
perfect (there is no electron's reflection  along the gluing
line), then the e--e interaction in the case is described by the
Luttinger model with four-component fermions. One can make sure
that, in this case, the cooperons will be eight-component
correlated complexes.
\par
It remains to discuss the experimental possibility to confirm the
form of the ground state in one-dimensional superconducting
channels. We have just shown that the distinctive feature of the
superconducting ballistic channels is the electroneutrality of the
ground state. This means that in the one-dimensional case only,
simultaneous addition of two Cooper pairs  is possible. (For the
case of the single-component fermions, it is $\hat a^{\dag}_R \hat
a^{\dag}_L$  and $\hat b^{\dag}_R \hat b^{\dag}_L$). Therefore, we
believe that the effects associated with coherent tunneling of
cooperons are promising for researches.
\par

\section{Conclusions}
In the paper, it was shown that, at a low temperature (less than $
2\pi v_c / L $), a Luttinger liquid with attraction between $
n$-component electrons is a system with a macroscopically large
number (i.e., increasing with increasing channel length) of the
Bose-like particles (cooperons) in one state. These correlated
complexes consist of $ 2n $ electrons (or holes). It is the
maximum possible number of fermions of which the existence at one
point is allowed by the Pauli principle. Although the wave
function of each correlated complex is non-invariant with respect
to the gauge transformation, the ground-state wave function turns
out to be invariant with respect to this transformation.
Therefore, the symmetry of the ground state wave function
coincides with the symmetry of the Hamiltonian. A global
spontaneous violation of gauge invariance does not occur in the
system. It takes place because the ground state degeneration
temperature in the case of one-dimensional superconductivity turns
out to be higher than the temperature of the phase transition to
the normal state. As a consequence of this fact, it is impossible
to introduce an order parameter in one-dimensional
superconductivity. Nevertheless, the presence of the macroscopic
number of cooperons in the ground state ensures a long-range order
with the suppression of cooperon-impurity scattering and, as the
consequence, to the absence of relaxation in the permanent
supercurrent in a 1-D channel. Also, the absence of charge
degeneration of the ground state (see Equations~(\ref{gr12f}) and
(\ref{gr12mn})) permits adding cooperons only by pairs that
conserve the condensate charge.


\vspace{6pt}




\par
This research received no external funding.
\par

\acknowledgments{We are grateful V.Yu. Petrov for the very useful discussions.}

\par
The authors declare no conflict of interest.







\begin{thebibliography}{999}
\bibitem{rev2}
Often the Luttinger model is called exactly solvable. This is true, but only in the sense that one can calculate any $ n-$particles Green function. Calculating the ground state wave functions is much more complicated procedure. We will see that it can be make analytically only in the limit of very strong e-e interactions and one-component fermions.
\end{thebibliography}
\end{document}